# Estimating city-level travel patterns using street imagery: a case study of using Google Street View in Britain

Rahul Goel, Leandro M. T. Garcia, Anna Goodman, Rob Johnson, Rachel Aldred, Manoradhan Murugesan, Soren Brage, Kavi Bhalla, James Woodcock

## Abstract

**Background:** Street imagery is a promising big data source providing current and historical images in more than 100 countries. Previous studies used this data to audit built environment features. Here we explore a novel application, using Google Street View (GSV) to predict travel patterns at the city level.

**Methods:** We sampled 34 cities in Great Britain. In each city, we accessed GSV images from 1000 random locations from years overlapping with the 2011 Census and the 2011-2013 Active People Survey (APS). We manually annotated images into seven categories of road users. We developed regression models with the counts of images of road users as predictors. Outcomes included Census-reported commute shares of four modes (walking plus public transport, cycling, motorcycle, and car), and APS-reported past-month participation in walking and cycling.

**Results:** In bivariate analyses, we found high correlations between GSV counts of cyclists (GSV-cyclists) and cycle commute mode share (r=0.92) and past-month cycling (r=0.90). Likewise, GSV-pedestrians was moderately correlated with past-month walking for transport (r=0.46), GSV-motorcycles was moderately correlated with commute share of motorcycles (r=0.44), and GSV-buses was highly correlated with commute share of walking plus public transport (r=0.81). GSV-car was not correlated with car commute mode share (r=-0.12). However, in multivariable regression models, all mode shares were predicted well. Cross-validation analyses showed good prediction performance for all the outcomes except past-month walking.

**Conclusions:** Street imagery is a promising new big data source to predict urban mobility patterns. Further testing across multiple settings is warranted both for cross-sectional and longitudinal assessments.

## Introduction

Urban mobility data is crucial to understand travel patterns, to plan and evaluate policies and interventions, and to analyse their social, health and environmental impacts (de Sá et al., 2017, Banister, 2008). However, gathering accurate, timely, and representative mobility data is not a trivial task. For instance, household travel surveys provide good data on personal travel patterns, but they are resource-intensive, often available only at the national or regional level, conducted infrequently, and aimed at long-term transport planning. Some countries include questions on travel behaviour in their census, but typically only cover commuting to work, and conducted infrequently (usually once per decade). Motor vehicle, cycle or pedestrian counts – by human observers or sensors – are more common, but usually cover specific areas or junctions within cities and lack representativeness, and are therefore difficult to compare across cities.

In this context, it is imperative to develop innovative, comparable, and cost-effective approaches to estimate walking, cycling and other travel patterns in cities. Recent studies have used big data sources to estimate travel metrics that are limited in detail but are available at a much larger scale than is possible through most surveys. For instance, Calabrese et al. (Calabrese et al., 2013) used cell phone data of a million users to estimate travel distance and number of trips in the Boston area. Althoff et al. (2017) used accelerometry data of more than 700,000 smartphone users across 111 countries to measure physical activity in terms of number of steps taken while carrying the device. There are similar studies using anonymised cell phone data to estimate the movement of individuals (Gonzalez et al., 2008, Kung et al., 2014).

Jestico et al. (2016) used crowdsourced GPS data of a cycle fitness application and found linear relationship with hourly cycle counts at multiple locations. These studies have also shown that with big data sources the analysis can often include more than just one country and thus has a global reach. However, the above studies have limitations in terms of their inability to identify the mode of travel, except when the data collection mechanism is specific to the method of travel (Althoff et al., 2017, Jestico et al., 2016). Studies which reported detection of travel modes used GPS tracks and are currently few in number. Moreover, the scope of these studies was limited to validation of the detection algorithm using a small dataset (Reddy et al., 2010, Nitsche et al., 2014). Studies using GPS tracks are limited to smartphones users, and therefore, also likely to suffer from the bias associated with the ownership of such devices (Smith, 2013). Further, their global coverage is limited by their use of datasets from specific mobile phone companies or smartphone applications.

Street imagery is a novel data source that provides visual information of the streets in the form of panoramic images. This includes static built environment features, people on the streets as pedestrians or cyclists, and vehicles. There are several providers of street imagery—Google Street View (GSV), Bing StreetSide, and four providers specific to China (Baidu, Netease, Amap, and Tencent (Long and Liu, 2017)). Among these, GSV is the largest, with full or partial coverage in 102 countries (although it does not include China and India) according to information available in October 2017 (Wikipedia, 2017, Google, 2017). Together these providers cover more than 50% of the global population spread across most of the world regions except large parts of Africa and central Asia (Google, 2017, Long and Liu, 2017, Wikipedia, 2017). GSV has been widely tested as a built environment audit tool (Badland et al., 2010, Rundle et al., 2011, Vanwolleghem et al., 2016). In China, Tencent has been used to create a street greenery index across 254 cities (Long and Liu, 2017).

However, what is less tested is the potential of street imagery resources to offer a consistent, scalable, and efficient resource to obtain travel data. We are aware of only one study that tested the utility of street imageries using GSV to estimate travel patterns, and this was limited to pedestrian volumes across three cities (Yin et al., 2015). This study was conducted in the US, and found a good relationship between manual pedestrian counts at 200 street segments and the counts detected using GSV. These results suggest further work to explore the utility of this resource including multiple settings and multiple road users, and at larger scale. Although images of people are pixelated, GSV has the potential to provide demographic information on pedestrians and cyclists, e.g. gender. This may be of particular interest for cycling which in some settings has substantial gender inequality (Pucher et al., 2011, Aldred et al., 2016).

In this study, we aimed to investigate the potential of GSV to estimate city-level travel patterns for active and motorised modes, by looking at relations between GSV images and routine surveillance data sources, and developing prediction models in a sample of cities in Great Britain. In addition, we did preliminary testing of the ability to predict the gender split of cyclists.

# Materials and methods

## Sample of cities

We used primary urban areas (PUAs) as our unit of analysis. PUAs are combinations of local authorities covering a contiguous built-up area, and can be considered equivalent to a city-region. As of 2016, there were 63 PUAs in Great Britain—55 in England, three in Wales, and four in Scotland (www.centreforcities.org/puas/). We selected a sample of 34 PUAs from England (n=29), Wales (n=3) and Scotland (n=2). Of these, 25 were randomly selected from the three countries, and a further nine were purposively selected to cover all the UK Biobank centres (Biobank, 2014) except London. The 34 sampled PUAs comprise 75 local authorities (one to nine local authorities per PUA). The mean 2011 Census population of sampled PUAs was 499,000 (standard deviation = 532,000; data accessed from www.nomisweb.co.uk/). In the rest of the paper, we refer to these PUAs as 'cities'.

# Google Street View

GSV imagery consists of a continuous series of 360-deree panoramas. Each panorama is developed by stitching together multiple overlapping images and is unique to the location and the time when its images were captured.

## Accessing images from Google Street View

The process of obtaining an image for a given location and in a given direction is automated through the use of Application Programming Interface (API). The GSV API uses geographic location or the unique panorama ID as one of the inputs. The GSV metadata API uses geographic location as an input, and reports year and month (yyyy-mm format) of the corresponding panorama and its unique ID. Using a geographic location, the API only provides the latest image taken and the metadata corresponding to it. We observed that the year reported in the metadata of the latest images varied spatially within and across the cities.

Given that cycling levels have changed in many local authorities over the last decade (Aldred et al., 2017), the images from GSV should be as temporally close to the comparison data as possible; here Census and Active People Survey (APS) (more details in the next sections). The latest Census was conducted in 2011 and APS has been conducted every year since 2005. Therefore, a common year of 2011 is preferred or, the range from 2010 through 2012.

In order to control the year of the images through the API, we used an open source Python package called 'streetview', which makes use of GSV's JavaScript API to access older images (Letchford, 2017). This package uses geographic location as an input and reports the metadata of the most recent as well as historical panoramas available within 5 meters of the location. Using this information, panorama IDs of the period of interest can be selected. Next, we use panorama ID in the API calls instead of the geographic location, thus controlling for the year when images were captured.

## Data collection using Google Street View

To generate data on travel patterns, we identified and recorded the number of different types of road users/vehicles appearing in the GSV images. To define categories of road users/vehicles we reviewed images in multiple British cities. Most transport modes could be classified as pedestrians, cyclists, motorcyclists, cars/taxis (henceforth 'cars'), buses, and vans/trucks.

The observed motor vehicles in GSV images consisted of both moving and parked. This distinction, however, is not clear in many images. We also observed many images with parked cycles, for which the distinction is almost always clear. Even when a cycle is partially visible without the cyclist, the location of the cycle (whether on the carriageway or the side of the road) can be used to differentiate between a parked and a moving cycle. Therefore we included two categories for cycles—cyclists and parked cycles. We did not make an equivalent distinction for pedestrians or any other type of vehicle. In total, we included seven categories—pedestrians, cyclists, parked cycles, cars, motorcycles, buses, and vans/trucks. The last category includes all the vehicles that we classified as for commercial use.

To facilitate observation of the images and recording of the data, we developed a webpage where the images from the API were accessed using the URL in real time. This eliminated the need to download images (prohibited by the terms of use of the API). Alongside the image, we included a questionnaire to record the presence of different road users. In the questionnaire, we included the seven categories of road users, and within each category, we included four options of counts: 0, 1–3, 4–6, and >6. A snapshot of the webpage is shown in Fig 1.

**Fig 1. Snapshot of the web-based questionnaire.**

We divided the data collection among four research assistants. We trained them to detect and identify different road user and vehicle categories in our questionnaire. Before the data collection for the present study, we conducted a small sub-study and assessed inter-rater agreement evaluation procedures. Further, during the data collection, we verified the quality of the classification by reassessing a sample of images of every city. The research assistants recorded their response by selecting a radio button for each road user category.

*Gender identification*

After completing the observations of the roads users, we developed another questionnaire in which we included only those images that include a cyclist, from across all the cities. The images were displayed on the webpage, while we used a pen-and-paper method to record the gender of cyclists from the images. The objective of this effort is to compare the gender ratios from GSV images with those reported in Census and APS. Both the datasets include respondents 16 years or older, therefore, we excluded child cyclists in gender identification. Data collection was conducted by two research assistants, one male and one female. They were not aware of the city to which the images belonged. In cases of disagreement on gender, an agreement was reached with mutual consent. In case of no agreement, we excluded that cyclist from the evaluation of gender ratios.

## Sampling of locations

GSV images are available for all the street locations of the cities. For our data collection, we sampled locations in each city, for which we used road network data from geographic information system (GIS). The GIS network is a collection of links, and the two ends of a link usually represent an intersection. The mean is 22,760 links in the cities with a maximum of 124,600 in Manchester. We used a multistage sampling process for the locations. In the first stage, we sample one point on each link. For this, we used 'sample.line' function of 'sp' package in R. In the second stage, we sample a small set of points from those sampled in the first stage. Next, we obtained historic metadata for these points. In the third stage, from the points for which metadata is available, we sample points that correspond to the 3-year period common to both the Census and the APS (2010-2012).

To determine the adequate sample size for the second and the third stage, we used Cambridge as a test case. We sampled 2000 random locations in the city. To cover the 360-degree view at each location, we obtained four images corresponding to the four headings (0, 90, 180, and 270°) where GSV panoramas were available. Next, we completed observations for the images using the questionnaire. We determined an adequate sample size of 1000 images. With this sample size, the number of observations in different categories of the questionnaire, expressed as a ratio of total images observed, stabilised.

With four headings at each location, we found that the field of view of two adjacent images overlapped; therefore, the same road user could appear in two images leading to double counts. We restricted the headings to two opposite directions—0 and 180°. We found that the selection of pair of opposite directions (0–180° or 90–270°) had no impact on the results and the proportions still stabilised around 1000 images.

Therefore, for the third stage of sampling we aimed at a sample size of 1000 locations, or 2000 images, in each city. This is twice the number of images at which we observed stabilisation of proportions for Cambridge. To achieve 1000 useful locations, we sampled 2000 locations in the second stage of sampling. We oversampled by 100% as many points will be eliminated because of (a) absence of any GSV panorama at the location or (b) absence of a panorama corresponding to the years of the current analysis (2010–2012).

From the second stage of sampling, out of 68000 random locations (2000 × 34 cities), 94% had at least one panorama. The years of the panoramas spanned over a period of 10 years (2008 – 2017). For the 3-year period of our interest, 2010 images are available for only 9% of the locations, while 83% locations have 2011 or 2012 images or both. For consistency across the cities, we restricted our sampling to 2011–2012 as all the cities have images corresponding to this period.

We first selected all the panorama IDs corresponding to 2011. With this selection, if the number of images were less than 1000, we selected the rest from the locations with 2012 images. Similarly, other years were included when inclusion of 2011 or 2012 images did not total to 1000 locations. The preference order for this selection was 2010>2009>2008. No locations were repeated in this process. In the final sample of 34000 locations (1000 × 34 cities), 0.4% are 2008 images, 9.5% are 2009 images, 0.3% are 2010 images, 19.5% are 2011 images and 70% are 2012 images.

## Google Street View observations

In preliminary analyses, we found that the distribution of images with cars among the count categories is similar across the cities (Coefficient of variation (CV): 0.1 for 1-3; 0.12 for 4-6; 0.33 for >6). Therefore, using the categorical variable of car does not add information as opposed to using the total number of images with cars. In case of all other modes, we found that in most cities the two categories of counts with four or more road users did not have any observations. Given the sparseness of this matrix, we decided to discard the count categories, and used only the total number of images with observations for further analysis. We refer to the GSV counts as **GSV Walk**, **GSV Cycle**, **GSV P–Cycle** (for parked cycles), **GSV Bus**, **GSV Car** and **GSV MC** (for motorcycles). We do not use observations of vans/trucks for further analysis.

We also calculated the proportion of images in each city by month and we refer to these as monthly proportions (**GSV Jan**, **GSV Feb**, and so on). We also categorised the months into four seasons—spring (March to May), summer (June to August), autumn (September to November), and winter (December to February).

## Census and Active People Survey

Census in all three countries within Britain is conducted every 10 years, most recently in 2011. In England and Wales, for those who had a job during the last one week and aged 16 years or older (16+), the census includes the following question—"How do you usually travel to work? Tick one box only. Tick the box for the longest part, by distance, of your usual journey to work?". In Scotland, the Census is done independently of England and Wales. The question on travel is applicable to those currently in employment as well as those studying. Census of Scotland also reports data harmonised for the UK in which full-time students are excluded. We accessed sex- and mode-specific data for both Censuses (harmonised data for Scotland) at local authority level for respondents aged 16 to 74.

Active People Survey (APS) was an annual cross-sectional survey which has been conducted in England (not Wales or Scotland) every year since 2005, except 2006-07. Each round of the survey is conducted over a period of one year starting from October. The sampling frame covers all individuals aged 16+ living in England with no upper age limit [England]. The sample size is a minimum of 500 individuals per local authority, with some local authorities choosing to boost their sample size in some years. The survey is conducted using computer-assisted telephone interviewing technique. In every sampled household, only one eligible person is interviewed.

APS reports prevalence of walking and cycling, and frequency and duration of the activities (number of days over the past month and length of time per usual day). This information can be further classified for utility (transport-related) as well as non-utility purpose. Survey data is weighted to be representative of the 16+ population of each reporting geography. For this analysis, we used a combined data of the period 2011–2012 and 2012–2013, which overlapped with the years of 90% of the images.

We aggregated both Census and APS data reported at local authority level to respective PUA levels. Census variables are available for all the 34 sampled cities, while APS variables are available only in 29 English cities. We expressed Census travel to work data as commute mode shares—percent of all workers travelling to work by a given mode. The denominator excluded those who reported working from home. The mode shares have been classified into walk, cycle, cars (combination of car driving, car passenger and taxi), bus, and PT+Walk (combination of walk, bus, underground and train). Walking as commuting main mode in Britain is relatively uncommon compared with overall walking levels (Goodman, 2013). Since PT often includes a walking stage (Service, 2017), we created the combined variable of PT+Walk. We refer to the mode shares as **Census Walk**, **Census Cycle**, **Census Bus**, **Census PT+Walk** (public transport and walk combined), **Census MC** (for motorcycles), and **Census Car**.

We used 12 variables from APS—three measures classified into four sub-groups. The three measures are prevalence (percent of respondents reporting to have done any walking or cycling in the past four weeks), the average number of days per week of walking or cycling among the respondents who reported any walking or cycling, and the average duration of walking or cycling per day among the respondents who reported any walking or cycling. The four sub-groups are all-purpose cycling, all-purpose walking, utility cycling, and utility walking. The utility part (for transport purpose) of walking and cycling reported in APS is calculated indirectly as the difference in walking (or cycling) for all purposes and walking (or cycling) for recreation or health purpose. The variables are named as **APS Prev All Cycle**, **APS Duration All Cycle** and **APS Days All Cycle**, and similarly for other nine measures, with 'All' replaced by 'Utly' for utility walking/cycling and 'Cycle' replaced by 'Walk'.

We calculated cycling gender ratios for all-purpose cycling and utility cycling from APS and commuter cycling from Census. The ratio was calculated as prevalence (or mode share) of cycling among males to the prevalence (or mode share) among females. We refer to these ratios as **APS Prev All Cycle M/F**, **APS Prev Utly Cycle M/F** and **Census Cycle M/F**.

## Statistical analyses

### Regression models

To look at simple relations we developed a Pearson correlation matrix to understand the association of Census and APS measures with GSV outputs. There is variation in what time of year images were captured across the cities. Seasonal variation of physical activity levels has been reported for many temperate settings in the world (Tucker and Gilliland, 2007). Thus, to investigate this potential source of bias, we developed a correlation matrix between the proportions of GSV images of different road users in each month.

Some modes have strong correlations with their respective GSV counts, while some are weakly correlated with their GSV counts and strongly or moderately correlated with GSV counts of other modes. Therefore we developed multivariable regression models in which Census and APS measures are predicted using GSV counts of multiple modes along with monthly proportions of images. Among these measures, mode shares and prevalence are proportions while number of days and duration of activity are continuous variables. We developed the regression models appropriate for the two variable types. Census mode shares of the four modes (walk plus public transport, cycle, motorcycles, and car) as well as four APS-reported prevalence measures (all-purpose and utility prevalence of walking and of cycling) range between 0 and 1. For these eight measures, we developed beta regression models. Beta distribution is defined on the interval 0 to 1 and is therefore appropriate to model proportions (Ferrari and Cribari-Neto, 2004). We used 'betareg' package in R to develop beta regression models (Cribari-Neto and Zeileis, 2009).

Among the continuous variables, we found that the distribution of cycling measures (days and duration) were right skewed due to the three outlier cities of York, Oxford and Cambridge. Therefore, we used a robust linear regression model which uses M-estimators (unlike ordinary least square (OLS) method in linear regression) and is robust to outliers. We used 'rlm' function in R to develop the robust regression models. In case of walking measures, we used linear regression models using OLS. The exhaustive set of explanatory variables for all the models include six mode-specific observations from GSV (pedestrians, cyclists, parked cycles, motorcyclists, buses, and cars) and 11 month-specific proportion of images.

To test the performance of the models, we used leave-one-out cross validation (LOOCV). In this method, one data point is excluded (test set), while the model is developed using all the other data points (training set). Using the model, the value of the test data point is predicted and this process is repeated for the whole dataset. Given a small dataset (n=34 for Census and n=29 for APS), LOOCV is an appropriate method of cross validation since it retains nearly all data points in its training dataset, while still testing against the bias any outlier may bring.

In order to select the set of explanatory variables to be included in the model, we calculated the prediction residual sum of squares (PRESS)(Allen, 1974), which is defined as the sum of squared difference between observed and predicted values. While PRESS statistic can be used for variable selection, it is obtained using squares of errors. Therefore, we also calculated mean and median of the absolute differences between the predicted and observed values. We refer to these as mean (MAE) and median absolute errors (MDAE), respectively. In order to compare prediction performance across different measures, we used standardised residuals as recommended by Espinheira et al. (Espinheira et al., 2008) for beta regression models and calculated in 'betareg' package by default. We present mean (MSE) and median (MDSE) standardised errors.

For a given outcome variable, we developed a maximally controlled model using all the data points. In this model, we included all the GSV outputs except those with near-zero Pearson correlation with the outcome variable (Fig 2), and also included all the months that were correlated with GSV outputs. Next, using the LOOCV method, we calculated the PRESS statistic for the model. Then we sequentially removed the variables with low statistical significance if the removal also resulted in any reduction of PRESS statistic. We carried out this process until we selected the set of variables that minimised the PRESS statistic.

The final models for commute shares of walk plus public transport, cycle, motorcycle, and car are referred to as Model 1, 2, 3, and 4, respectively. For mode share of cycle (Census Cycle; model 2), the explanatory variable GSV Cycle is transformed as its square root. With the original form of GSV Cycle, the relationship was highly influenced by the outlying data point of Cambridge. We selected the square root relationship by analysing the scatterplot between the observed and predicted values (through LOOCV method) using GSV cycle in its original form. The scatterplot showed that the predicted values were related to observed values through a square function.

For all-purpose and utility cycling prevalence, final models are 5 and 6, respectively, and for all-purpose and utility walking, final models are 7 and 8, respectively. For average days of utility cycling and of utility walking, final models are 9 and 10, respectively. Additionally, we developed four linear regression models for average days of all-purpose cycling, average days of all-purpose walking, average duration of utility cycling, and average duration of all-purpose cycling. For the two other walking-related outcomes, average duration of utility and all-purpose walking, we do not present the regression models as these outcomes showed only weak correlations with the all the predictor variables.

We also present scatterplots of the observed and predicted values for all the 10 models. These plots also include a *y=x* line as a reference to compare the error in predicted values. In a scenario, when the predicted values are close to the observed ones, all the points will lie close to *y=x* line. The scatter of the points around *y=x* line also indicates the bias in the model. For instance, if for a part of the scatterplot, all the points lie below or above the *y=x* line, this indicates that the model is biased for that range of data points.

# Gender split of cyclists

We detected gender of cyclists in the 29 English cities. For England, gender equality of cycling has been reported to increase with the level of cycling (Aldred et al., 2016). We sorted the cities in the ascending order of the gender ratio of commuter cycling reported in Census (Census Cycle M/F). Next, we divided the cities into 4 groups such that the gender ratios within a group are similar and the total number of observations across the groups is also similar.

For each group, we calculated the gender ratios using the observed number of male and female cyclists. Further, for each group, we calculated the average gender ratios from Census and APS. The measure of gender ratio from Census is Census Cycle M/F, and two measures from APS are APS Prev All Cycle M/F and APS Prev Utly Cycle M/F. We calculated the average by weighing average ratio in each city with the number of observations of that city.

**Table 1. Descriptive statistics.**

|  | Mean | Std. Deviation | Minimum | 25th Percentile | Median | 75th Percentile | Maximum |
|---|---|---|---|---|---|---|---|
| Population | 498,758 | 532,382 | 107,053 | 211,826 | 339,864 | 578,604 | 2,422,818 |
| Number of local authorities per PUA | 2.2 | 1.7 | 1 | 1 | 5 | 2.8 | 9 |
| **Measures from Census (travel to work)** | | | | | | | |
| Census Walk (%) | 12.7 | 3.3 | 8.7 | 10.1 | 11.7 | 15.1 | 19.8 |
| Census Cycle (%) | 4.8 | 6.1 | 1.1 | 1.8 | 2.7 | 4.8 | 32.5 |
| Census MC (%) | 0.8 | 0.3 | 0.2 | 0.6 | 0.7 | 1.0 | 1.4 |
| Census Bus (%) | 10.6 | 5.0 | 5.0 | 7.2 | 8.3 | 13.6 | 28.6 |
| Census PT+Walk (%) | 28.0 | 7.6 | 16.2 | 23.2 | 27.9 | 30.7 | 49.4 |
| Census Car (%) | 65.9 | 10.8 | 36.8 | 62.6 | 67.4 | 72.5 | 81.3 |
| Census Cycle M/F (ratio) | 2.49 | 1.05 | 1.02 | 1.78 | 2.33 | 2.98 | 5.61 |
| **Measures from APS** | | | | | | | |
| APS Prev All Cycle (%) | 16.4 | 9.4 | 8.1 | 11.7 | 13.9 | 17.9 | 53.4 |
| APS Days All Cycle (per week) | 0.42 | 0.40 | 0.15 | 0.24 | 0.29 | 0.40 | 2.11 |
| APS Duration All Cycle (h per day) | 0.40 | 0.27 | 0.19 | 0.28 | 0.32 | 0.39 | 1.59 |
| APS Prev Utly Cycle (%) | 8.9 | 9.5 | 1.7 | 4.3 | 5.7 | 8.3 | 47.5 |
| APS Days Utly Cycle (per week) | 0.27 | 0.36 | 0.05 | 0.10 | 0.14 | 0.23 | 1.80 |
| APS Duration Utly Cycle (h per day) | 0.21 | 0.24 | 0.04 | 0.11 | 0.14 | 0.18 | 1.29 |
| APS Prev All Walk (%) | 85.2 | 2.4 | 79.7 | 83.7 | 85.3 | 86.9 | 90.8 |
| APS Days All Walk (per week) | 3.56 | 0.18 | 3.27 | 3.42 | 3.53 | 3.67 | 4.13 |
| APS Duration All Walk (h per day) | 4.14 | 0.37 | 3.49 | 3.93 | 4.19 | 4.38 | 5.04 |
| APS Prev Utly Walk (%) | 58.4 | 5.4 | 47.6 | 53.8 | 58.8 | 61.6 | 73.5 |
| APS Days Utly Walk (per week) | 2.06 | 0.24 | 1.66 | 1.87 | 2.03 | 2.17 | 2.54 |
| APS Duration Utly Walk (h per day) | 2.65 | 0.29 | 2.09 | 2.49 | 2.70 | 2.81 | 3.23 |
| APS Prev All Cycle M/F (ratio) | 2.36 | 0.61 | 1.16 | 2.06 | 2.34 | 2.71 | 3.70 |
| APS Prev Utly Cycle M/F (ratio) | 3.66 | 2.42 | 1.18 | 2.22 | 2.92 | 4.22 | 11.29 |
| **Measures from GSV (counts of images)** | | | | | | | |
| GSV Cycle | 18 | 20 | 3 | 6 | 14 | 19 | 94 |
| GSV P-Cycle | 16 | 32 | 0 | 3 | 6 | 10 | 132 |
| GSV Walk | 209 | 56 | 138 | 170 | 192 | 238 | 371 |
| GSV Car | 1438 | 115 | 1111 | 1372 | 1430 | 1532 | 1620 |
| GSV Bus | 27 | 14 | 11 | 17 | 23 | 34 | 74 |
| GSV MC | 11 | 7 | 1 | 6 | 10 | 14 | 42 |
| **Seasonal distribution of GSV images** | | | | | | | |
| GSV Autumn (%) | 26.9 | 24.9 | 0 | 7 | 24.7 | 35.5 | 94.2 |
| GSV Spring (%) | 30.3 | 28.9 | 0 | 1.9 | 26.1 | 45.9 | 94.4 |
| GSV Summer (%) | 42.7 | 27.5 | 5.2 | 26.4 | 34.3 | 65.3 | 99.6 |
| GSV Winter (%) | 0.03 | 0.1 | 0 | 0 | 0 | 0 | 0.7 |

# RESULTS

**Table 1** presents the descriptive statistics of Census and APS variables at PUA level. From a total of 68000 images, we observed the following number of images with at least one of these road users— 7101 (pedestrians), 620 (cyclists), 534 (parked cycles), 48900 (cars), 366 (motorcycles), 925(bus) and 11184 (vans/trucks). These do not total to 68000 as one image can have more than one road user. There is large variation in the city-specific number of images for different road users. The highest number of images is for cars (average 1438 across 34 cities) followed by pedestrians (209), while the number is much lower for other road users, varying from an average of 11 for motorcycles to 27 for buses.

The variation in the number of images across the cities is the highest for cyclists and parked cycles (CV: 1.1 and 2, respectively), lowest for cars (0. 08), and in-between for other modes (Walk: 0.27, Bus: 0.52, motorcycles: 0.63). The average share of images from summer months (June to August) is the highest (43%) followed by autumn (September to November, 27%) and spring (March to May 30%), with almost no images from winter months. The distribution across the three seasons is not uniform across the cities. For instance, some cities have almost all their images from one season (maximum value of 94% to 100%).

Among the APS variables (Table 1), prevalence of all-purpose walking is on an average 5 times higher than all-purpose cycling. The prevalence of utility walking is proportionally even higher than its cycling counterpart (>6 times on an average). The divide between the two modes is similarly high in terms of duration and days. Among the Census variables, the commute share of walking is on an average 2.5 times higher than cycling.

# Correlation of GSV observations with Census and APS measures

Fig 2 presents correlation chart for Census and APS measures along with GSV observations. Figs 3 and 4 present scatterplots of GSV observations with measures of Census and APS for selected variables. Also shown are the $R^2$ values for a linear line fitted to the data.

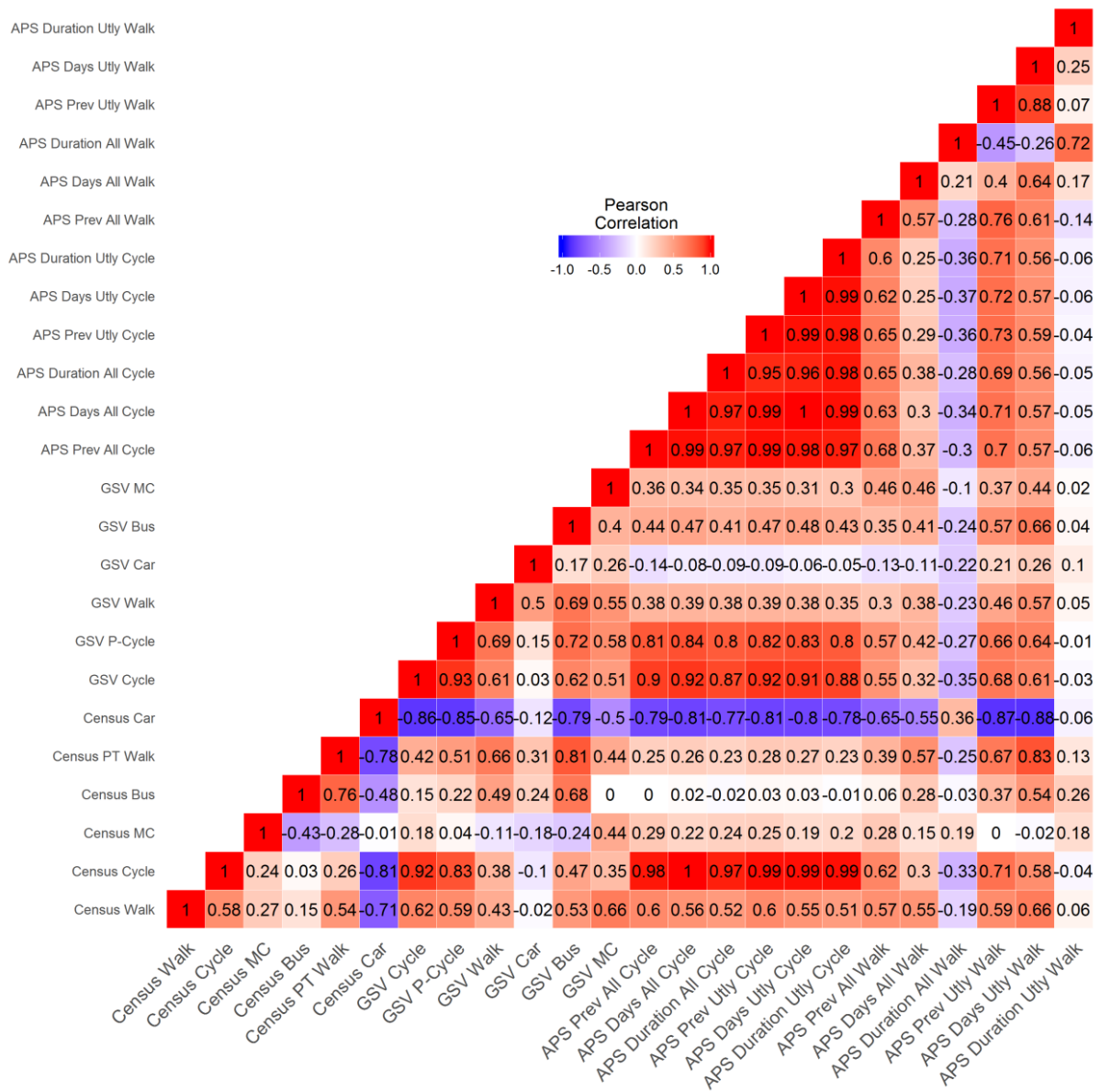

**Fig 2. Pearson correlation among GSV, APS and Census variables.**

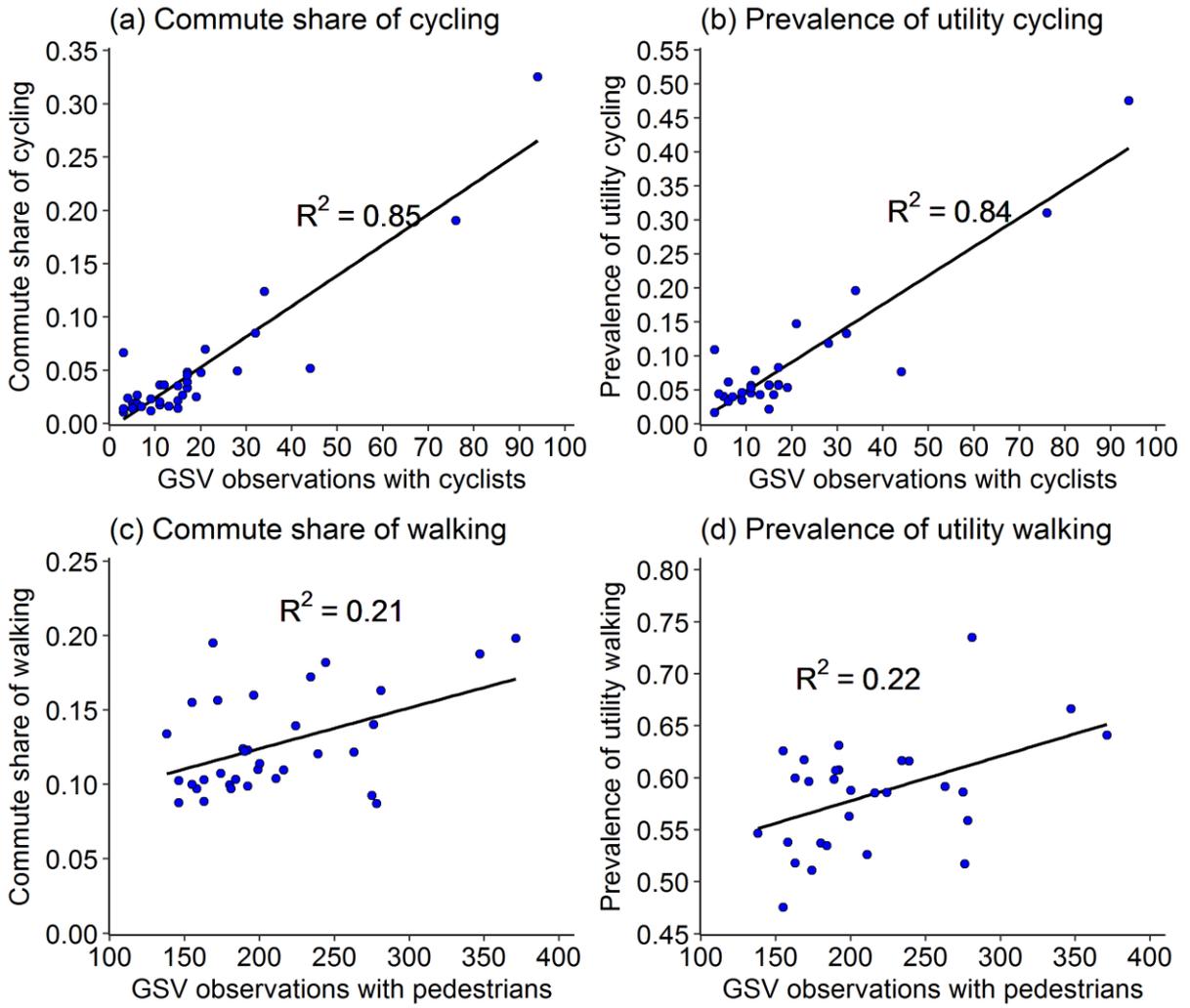

**Fig 3. Linear relationships of GSV observations with commute share and prevalence of walking and cycling.**

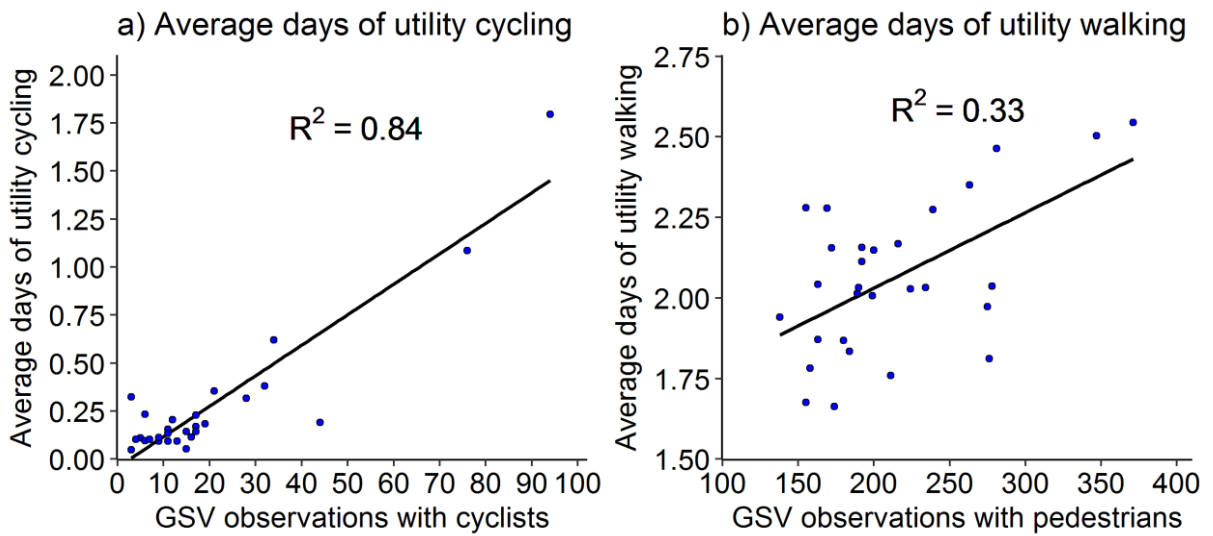

**Fig 4. Linear relationships of GSV observations with APS measures of walking and cycling.**

Among all the GSV measures, GSV Cycle has the highest correlation with its corresponding APS as well as Census variables. The correlation ranges from 0.87 to 0.92 with all the cycling-related measures of APS. GSV Cycle has equally high correlation (0.92) with commute share of cycle. High correlations (0.8 to 0.84) also exist between GSV Parked Cycles and cycling-related variables in Census and APS, though lower than GSV Cycle. In addition to cycle-related variables, GSV Cycle also has high positive correlation with commute share of walking (0.62) and a high negative correlation with commute share of cars (–0.86).

The correlations of GSV Walk with walking-related variables from APS and Census range from weak to moderate. Among APS variables, GSV Walk has moderate correlations with prevalence and number of days of all-purpose and utility walking. Within these, the correlation are higher for utility walking (0.46 to 0.57) than all-purpose walking (0.3 to 0.38). Correlations with duration-related variables are much weaker. Among Census variables, GSV Walk has a correlation of 0.43 with commute share of walk, and a correlation of 0.66 with the combined commute mode share of public transport and walk (Census PT Walk).

Among APS variables, GSV Bus has moderate to high correlations with prevalence and days of walking (0.37 to 0.66) and moderate correlations with all the measures of cycling (0.41 to 0.48). Among Census variables, GSV Bus has a high positive correlation with commute share of bus (0.68), even higher correlation with combined commute share of public transport and walk (0.81), and a high negative correlation (–0.79) with commute share of cars.

GSV Car has weak correlation with all the APS-related variables. Among Census variables, GSV Car has a weak negative correlation with commute share of car (–0.12). Further, it has a weak to moderate positive correlation with commute shares of buses (0.24) and combined walk and public transport share (0.31). It has almost no correlation with the commute shares of all the other modes. Among APS variables, GSV motorcycles has moderate correlations with prevalence and days of walking (0.37 to

0.46) and with all the measures of cycling (0.3 to 0.36). Among Census variables, GSV MC has a moderately high correlation with the commute share of motorcycles (0.44) as well as all the other modes—cars (–0.5), walking (0.66), cycling (0.35), and combined walking and public transport (0.44).

## Effect of seasonality of images

The proportion of images from one season have a large variation, from 0% to more than 90%, with a large standard deviation (Table 1). GSV observations of cars and pedestrians have negative correlations with spring months, and positive for all others, except negative correlation of pedestrians with June and November. GSV Cycle has a negative correlation with March, all of winter and most of autumn, and positive correlation with August and September. GSV MC has a positive correlation with summer months and negative for most other months.

## Regression models

Table 2 presents all the regression models (1 through 10) using all the data points in the dataset. The table also presents mean (MAE) and median of the absolute errors (MDAE). Figs 5-7 present the scatterplots of the observed and predicted values for the 10 models. GSV Cycle is a predictor across all the models except APS Days All Walk (model 10). GSV Bus is a predictor for all the mode share outcomes except that for cycle, in which case GSV Cycle is the only predictor.

**Table 2. Regression models.**

|  | Beta regression models |  |  |  |  |  |  |  | Linear regression |  |
|---|---|---|---|---|---|---|---|---|---|---|
|  | Model 1 | Model 2 | Model 3 | Model 4 | Model 5 | Model 6 | Model 7 | Model 8 | Model 9* | Model 10 |
|  | Census PT+Walk | Census Cycle | Census MC | Census Car | APS Prev All Cycl | APS Prev Utly Cycl | APS Prev All Walk | APS Prev Utly Walk | APS Days Utly Cycl | APS Days Utly Walk |
| (Intercept) | -1.356 | -4.877 | -4.759 | 1.223 | -1.995 | -3.662 | 1.571 | 0.192 | -0.033 | 1.562 |
| GSV Walk |  |  |  |  |  | -0.004 |  |  |  | 0.002 |
| GSV Cycle | -0.004 | 0.408^ | 0.006 | -0.01 | 0.026 | 0.467 | 0.007 | 0.008 | 0.015 |  |
| GSV MC |  |  | 0.027 |  |  |  |  |  |  |  |
| GSV Car |  |  |  |  |  |  |  |  |  |  |
| GSV Bus | 0.020 |  | -0.018 | -0.016 | -0.008 |  |  |  |  |  |
| GSV Feb |  |  |  |  |  |  |  | 0.553 |  |  |
| GSV Mar | -0.013 |  |  | 0.010 |  |  | 0.009 |  |  |  |
| MAE | 0.04 | 0.015 | 0.002 | 0.04 | 0.03 | 0.02 | 0.02 | 0.03 | 0.11 | 0.17 |
| MDAE | 0.03 | 0.008 | 0.001 | 0.04 | 0.02 | 0.01 | 0.01 | 0.02 | 0.07 | 0.16 |
| MSR | 1.076 | 0.867 | 1.415 | 1.089 | 0.921 | 0.965 | 1.38 | 0.908 | - | - |
| MDSR | 0.865 | 0.653 | 0.712 | 0.959 | 0.735 | 0.679 | 0.933 | 0.629 | - | - |

*Model 9: robust linear regression; ^ For square root of GSV Cycle.

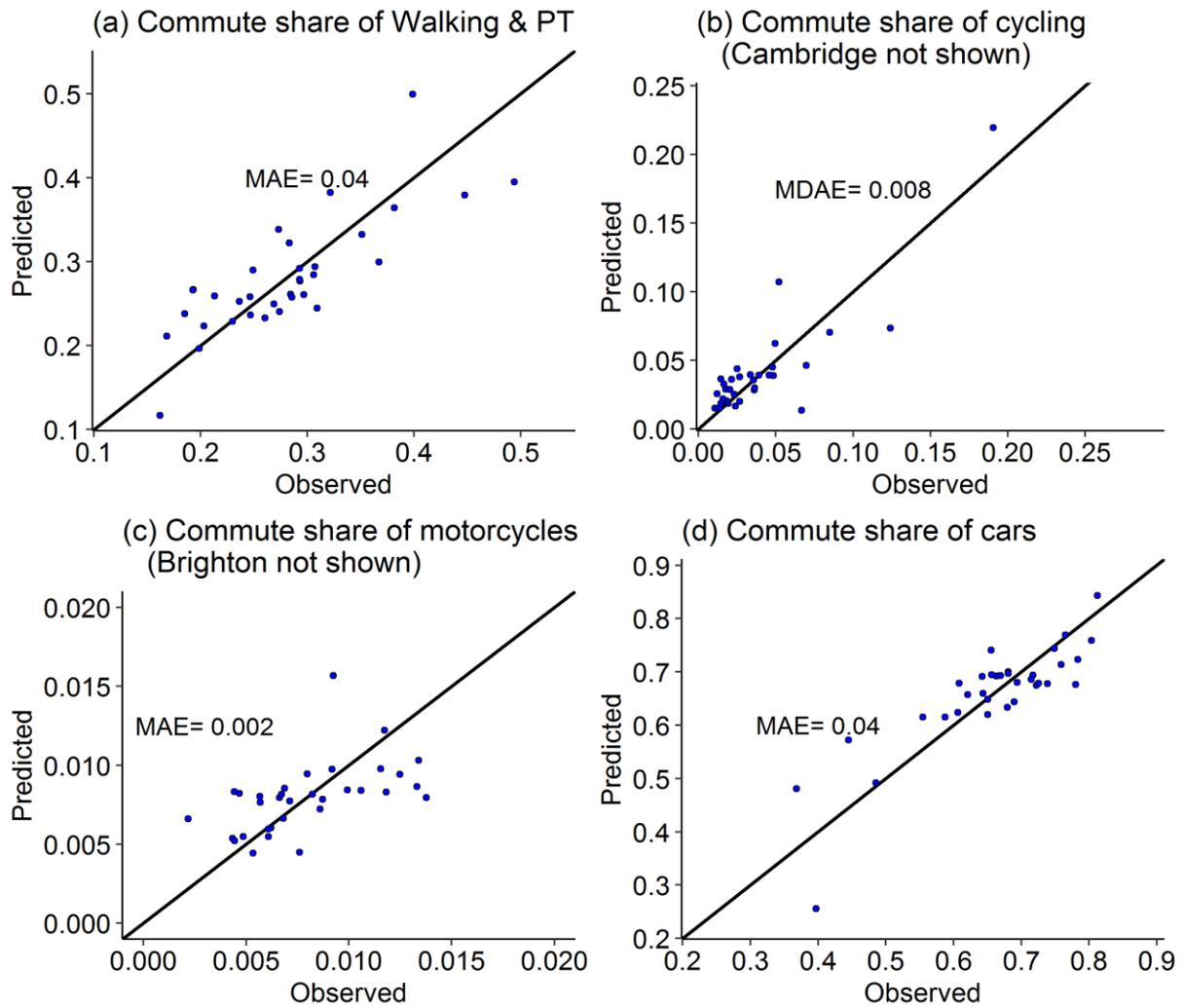

Fig 5. Observed and predicted mode shares using LOOCV with mean absolute error (MAE) and median absolute error (MDAE) for cycling measures (a: Model 1; b: Model 2; c: Model 3; d: Model 4)

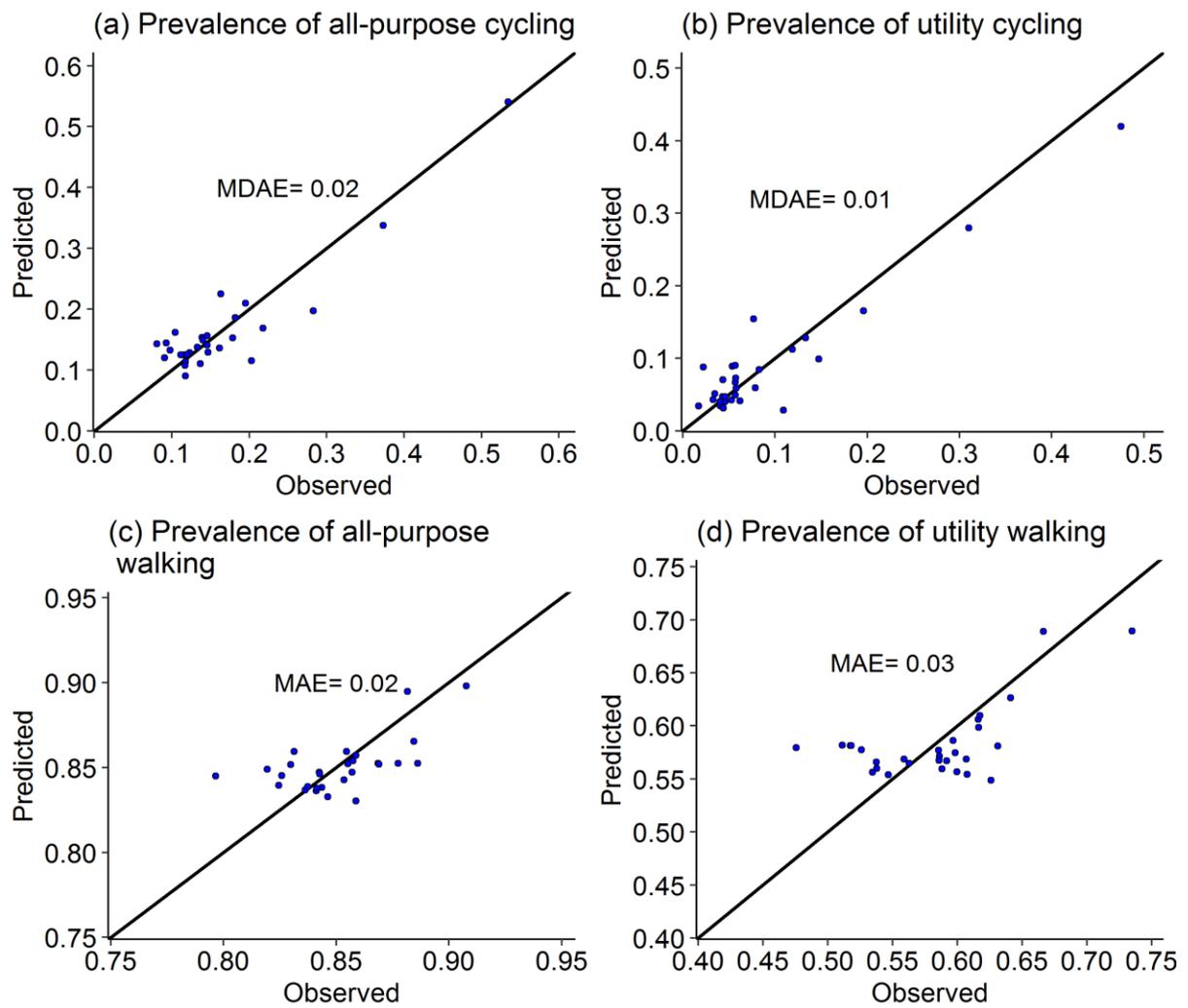

Fig 6. Observed and predicted prevalence measures using LOOCV with mean absolute error (MAE) and median absolute error (MDAE) for cycling measures (a: Model 5; b: Model 6; c: Model 7; d: Model 8)

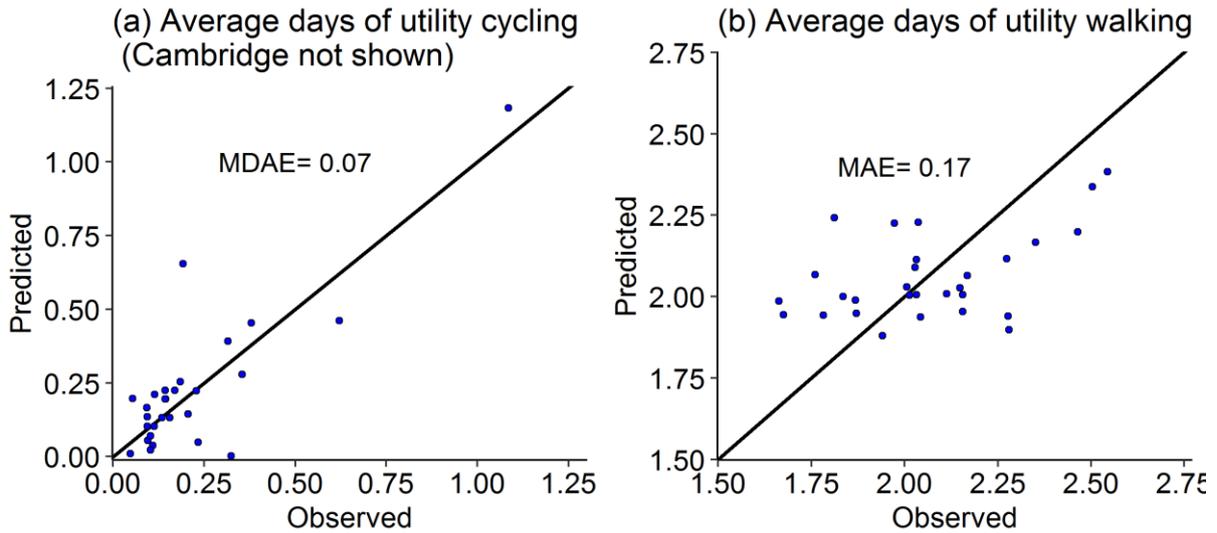

**Fig 7. Observed and predicted average number of days using LOOCV with mean absolute error (MAE) and median absolute error (MDAE) for cycling measure (a: Model 9; b: Model 10)**

## Commute mode shares

The scatterplots of the observed and predicted values of commute mode shares (Fig 5) show that the predicted values are scattered uniformly around and close to y=x line for all the modes. This shows consistent accuracy of prediction across the full range of data for all the modes. The MDAE are 3%, 0.9%, 0.1% and 4% for walk plus public transport, cycle, motorcycles, and cars, respectively. The corresponding median mode shares are 28%, 2.7%, 0.7% and 67%, respectively. Median standardised errors are the lowest for cycle (0.65) and the highest for cars (0.96).

The difference between MAE and MDAE is the highest for cycles (0.015 and 0.008, respectively) where the highest commute share in Cambridge (32.5%) is more than six times higher than the average across all the cities. The model predicts a mode share of 23.9% for Cambridge, which is an error of 8.6 percentage points, thus substantially increasing the mean error. This data point is not shown in the scatterplot (Fig 5a) for a better representation of the plot. In the scatterplot of motorcycles (Fig 5c), Brighton city is not shown with a predicted share of 2.4% compared to observed 0.9%.

## Walking and cycling measures in APS

The prediction results for APS prevalence measures show mixed results. The scatterplots of the observed and predicted values of prevalence (Fig 6) show that the predictions are more uniformly scattered around *y=x* line for cycling measures than for walking measures. MDAE for all-purpose cycling is 2% and that for utility cycling is 1.8% compared to their corresponding median observed values of 14% and 5.7%. The median standardised errors are 0.74 and 0.68, respectively.

Among walking prevalence, models for all-purpose as well as utility walking show poor prediction. The scatter around *y=x* line is not uniform, though the scatter is more uniform for all-purpose walking than utility walking. For all-purpose walking (Fig 6c) a large number of predicted values lie in a narrow range of 84% to 86% while their corresponding observed values range from 79% to 89%. Similarly, for utility walking (Fig 6d), a large number of predicted values lie in the narrow range of 55% to 60% while their corresponding observed values range from 47% to 64%. The median error for all-purpose walking is 1.3% and for utility walking is 2.4%, compared to their corresponding median values of 85% and 59%.

In Fig 7 we present the scatterplots of the average number of days of utility cycling and utility walking. The predicted values are uniformly scattered for cycling measure. The median error for cycling measure is 7% compared with median observed value of 14%. The scatter for walking measure is much more biased similar to walking prevalence discussed above. The predicted values lie between 1.8 to 2.1 corresponding to the observed range of 1.6 to 2.3.

## Gender split of cyclists

There are a total of 570 images with at least one cyclist. Among these gender was detected in 298 unique images (52% of 570) for 315 cyclists. Out of these 315, we identified 61 (19%) as children. The two researchers had no disagreement in identifying cyclists as children or non-children. For estimating the cycling gender ratio, we used 254 observations (315 minus 61), out of which 82 are females and 172 males. The overall ratio of males to females is 2.1 (172/82). The gender-specific observations are sparse at the city level. In many cases, for instance, there are no observations for females, and therefore, comparison of city-specific ratios is not possible.

Table 3 presents number of gender observations for all cities combined and for each group of cities. Group 1 has only Cambridge with 61 observations (male and female), group 2 has Oxford and York with a combined 60 observations, group 3 consists of 4 cities with 67 observations and lastly group 4 has 22 cities with a total of 76 observations. Note that from group 1 to 4, inequality in gender ratio increases. Table 3 also presents gender ratios observed from GSV images and those reported in the three measures from Census (commuting) and APS (all-purpose and utility cycling).

These results can be interpreted in two ways. In terms of direct comparison, GSV-based gender ratios are closest to APS prevalence of utility cycling (2.1 and 2.2). Group-specific comparisons show similar estimates for all the groups except 2. In terms of a relationship between group-specific estimates, GSV ratios have an approximate monotonic (non-linear) relationship with the estimates of all the three measures. Only in case of group 2, GSV estimates are higher than expected from a monotonic relationship.

**Table 3: Comparison of GSV, Census and APS estimates of gender split of cyclists.**

| Groups | Number of GSV | | Ratios of male to female | | | |
|---|---|---|---|---|---|---|
| | Females | Males | GSV Observations | Commute cycling share | APS Prevalence of all-purpose cycling | APS Prevalence of utility cycling |
| All | 82 | 172 | 2.10 | 1.63 | 1.82 | 2.21 |
| Group 1 | 28 | 33 | 1.18 | 1.02 | 1.28 | 1.31 |
| Group 2 | 18 | 42 | 2.33 | 1.16 | 1.33 | 1.29 |
| Group 3 | 18 | 39 | 2.17 | 1.51 | 2.07 | 1.81 |
| Group 4 | 18 | 58 | 3.22 | 2.60 | 2.45 | 3.98 |

# DISCUSSION

We investigated relationships between GSV observations of road users and their travel patterns reported in surveillance datasets (Census and APS) for 34 cities in Great Britain. We found that GSV observations are strong predictors of commute mode shares classified as walking plus public transport, cycle, motorcycles, and car. We also found that GSV observations are strong predictors of the past-month cycling activity levels reported by the APS. For cross-validation we used the 'leave-one-out' method and all the models performed well at prediction, except APS-reported measures of walking.

Among the GSV observations of different road users, cyclists are strong predictors across all the outcomes and buses are strong predictors for the commute mode shares. This may be because cyclists and buses had much higher variation across the cities than cars and pedestrians. For cyclists, we found promising initial results in the comparison between gender ratios observed in GSV images and the ratios reported in Census and APS. This analysis remained limited, however, by the small number of observations.

This is the first study that has investigated the use of a street imagery dataset to estimate a range of travel outcomes at the city level. We used street imagery from GSV, which has the largest coverage across the world. The explanatory variables used in the prediction models are observations from GSV images and the metadata (month of images); therefore, the methods presented are generic in nature and can be applied in any setting of the world where GSV is available. Excluding metadata, the availability of which may differ across other providers of street imagery, the methods reported in this study can be applied using any street imagery data source and not just GSV. Further, we used a small sample size of images in each city. This ensured that manual annotation of images was efficient as shown by stabilisation of relative observations for each mode after 1000 images. The strong relationships in the prediction models are further encouraging.

The application to test GSV is limited to what it can be calibrated to, i.e., what travel pattern outcome variables are available. In our study, the outcome variables are the commute mode shares reported by Census and past-month participation in walking and cycling reported by APS. Both Census and APS have their strengths as well as limitations, and these transfer to the relationships built with GSV. The Census is statistically powered at local authority level with its universal coverage of eligible respondents. However, it includes only a fraction of overall travel (commuting only) for a subgroup of population (16+ workers only). This information is further limited as it only refers to the usual, main mode of travel, which will particularly underestimate the use of walking in a multi-modal trip.

APS was conducted every year (now replaced by similar but on-line Active Lives Survey) and reports walking and cycling in all the activity domains for the whole of England. This makes APS a unique dataset in terms of its coverage of population and measures of active travel included in its questionnaire. APS, however, is conducted for a small sample of respondents, and is therefore weakly powered at local authority level. In addition, APS includes self-reported physical activity, which has its limitations due to respondent inaccuracies in reporting the frequency and duration of their physical activity.

The limitations in APS data may partly explain why we found good relationships for all the measures of cycling, but a poor relationship for APS-reported walking. On an average, 85% of all adults report some level of walking and this shows little variation across cities. Therefore, for an activity done frequently and for small distances, APS is less likely to capture the true differences across the populations using a self-reported questionnaire. Among all the walking metrics from APS, days of utility walking has the highest correlation with GSV observations of walking (r=0.57). With most variation across the cities, this measure may also be most sensitive to genuine variation in the level of walking. Given the relative rarity of cycling in Great Britain, the self-report questionnaire is more likely to distinguish across cycling levels in the populations by identifying individuals who do not cycle at all.

Similar to Census and APS, there are also limitations as well as strengths of the GSV observations we used as explanatory variables. Observations from GSV did not differentiate between non-transport street life and walking. This may bring additional differences between our explanatory variables and walking-related outcomes measures. However, this can also be identified as a strength of GSV. Given an increasing emphasis on developing streets to serve function of a 'place' and not just 'movement'(London, 2016), GSV can be used as a complementary data source to measure overall street life, since it captures precisely where a traveller is on the road network (but not where their trip started and ended), whereas the Census, for example, does the opposite.

Further, GSV API reports only a limited set of information in the metadata of the images (month and year). Time-of-day as well as day-of-week are not available in the API. Both these variables are likely to be significantly associated with GSV observations, and could introduce measurement error or even bias if the profile of images taken vary systematically across the cities. For a small number of images, we used shadows to approximately determine the time-of-day and found that most images are taken during inter-peak periods. While a limitation, this is also a strength as peak (commuting) data are better captured through censuses and counts. Therefore, in future, with the determination of time-of-day, GSV can be used as a complementary dataset to account for the travel activity that occurs during inter-peak periods.

The models presented in this study show that GSV images can be used as predictors of a variety of population-based measures of transport. We found that there is a strong direct relationship between cycle observations in GSV and cycling measures (modes share and past-month participation) with a high correlation (from 0.87 to 0.92). GSV counts of cyclists alone is a sufficient predictor of cycling measures. This means that, even without a prediction model, GSV observations can be used to develop a relative index of cycling levels across cities in Britain. Given that cycling remains a marginal mode of transport in many other countries, it seems plausible that this may generalise more widely. As such GSV offers a promising approach to provide a robust method for the continuous surveillance of cycling levels.

For modes other than cycling, the relationships are either multivariable or else the outcomes are best predicted by the GSV images of a different mode. These more complicated relationships might be expected to be more context specific. Future research should investigate how well predictive models work across different settings and if street imagery can be combined with other data to produce more generalisable models.

Given that the oldest images of GSV are available since 2007 and imagery in Great Britain has been updated almost every year, it would also be worthwhile to explore the potential to extend this method to the longitudinal evaluation of natural experiments. Future research should also use objective measures of travel physical activity to develop relationships with GSV observations, which are better suited than self-reported measures to capture differences across populations. For instance, the datasets such as UK Biobank reported objective measurements of physical activity levels in 100,000 participants across 22 cities in Great Britain **(Sudlow et al., 2015)**. Althoff et al. **(Althoff et al., 2017)** have reported objectively measured number of steps for multiple cities in the United States.

Improvement can also be made in the data collection of images. Using machine-learning based image recognitions programs, the sample size of images in each city can be increased multiple times. With this capability, With more stable estimates of observations, GSV observations are more likely to capture small differences across the populations, and therefore more fully capture the 'big data' opportunity that this resource represents for transport researchers and practitioners.

## Acknowledgements

We would like to thank Natascia Furlon, James Long, Callum Pinner and Eloise Waterton for their assistance in the data collection. We thank Anna Melachrou at MRC Epidemiology Unit for her assistance in developing the web-based questionnaire and Ali Abbas for his assistance with the API. We also thank Stephen Sharp at MRC Epidemiology Unit for his assistance in statistical analyses.